# PRE-TRAINING END-TO-END ASR MODELS WITH AUGMENTED SPEECH SAMPLES QUERIED BY TEXT

*Eric Sun, Jinyu Li, Jian Xue, Yifan Gong*

Microsoft

## ABSTRACT

In end-to-end automatic speech recognition system, one of the difficulties for language expansion is the limited paired speech and text training data. In this paper, we propose a novel method to generate augmented samples with unpaired speech feature segments and text data for model pre-training, which has the advantage of low cost without using additional speech data. When mixing 20,000 hours augmented speech data generated by our method with 12,500 hours original transcribed speech data for Italian Transformer transducer model pre-training, we achieve 8.7% relative word error rate reduction. The pre-trained model achieves similar performance as the model pre-trained with multilingual transcribed 75,000 hours raw speech data. When merging the augmented speech data with the multilingual data to pre-train a new model, we achieve even more relative word error rate reduction of 12.2% over the baseline, which further verifies the effectiveness of our method for speech data augmentation.

## 1. INTRODUCTION

Recently, in automatic speech recognition (ASR), pre-trained models are intensively explored with the self-supervised learning (SSL) fashion [1, 2, 3, 4] by mainly training an encoder from the huge amount of unlabeled data before applying a task-oriented supervised training with limited data. However, the gain of SSL keeps shrinking with the increased amount of fine-tuning labeled data [5, 6]. On the other hand, using transcribed data from resource-rich languages to do model pre-training has been a popular way in the industry for the language expansion [7, 8, 9, 10, 11, 12] of locales with medium amount of training data by using transfer learning technique which leverages the supervised data from multiple languages to pre-train a seed model and then fine-tune with the data from target languages [7, 8, 9, 10, 11, 12]. The drawback for transfer learning method is it needs lots of supervised data from multiple languages in order to better cover the acoustic conditions in the target languages. In addition, the time and computational cost is very high for pre-trained models due to the huge amount of supervised data from a variety of languages.

In this paper, we propose a new method which generates augmented speech data to train end-to-end (E2E) models by using text data to query ASR training corpus. We first generate the alignment between transcribed speech and the corresponding text-based grapheme letters [13]. We then extract speech feature segments and corresponding texts as pairs to build candidate libraries. Based on the texts that are from language model (LM) training data, we finally sample speech feature segments corresponding to underlying units of target texts from candidate libraries to concatenate them into new speech feature utterances. The main contributions of this paper are:

- A general framework for large-scale speech data augmentation by using text data to query ASR corpus to generate new speech feature utterances for E2E model pre-training. Compared to the Text-to-Speech (TTS) method [14, 15, 16, 17, 18] to generate speech data, the proposed method can cover more speaker and acoustic environment variations, which makes the pre-trained model more robust. It also has much lower cost since no TTS model is needed.
- A new method to build libraries of the speech feature segments for different levels of units from word, sentence pieces [19], to grapheme letters that can construct speech utterances for any texts to effectively avoid the issue of out of vocabulary (OOV) word.
- A significant reduction in the amount of data needed for the model pre-training compared to the method of transfer learning thereby directly resulting in cost savings.

For English, there are lots of training data available for the industry model development and pre-trained model is not usually applied. Therefore, as a case study, we use Italian with medium amount of training data as our target language which has 12,500 hours raw labeled speech data. We generate 20,000 hours speech utterances based on 9 million text sentences that are from Italian LM training data. A pre-trained model is built based on 37,500 hours raw speech data from both transcribed and augmented speech data. We show the pre-trained model is not able to obtain better accuracy than the baseline model trained with only original transcribed data because the augmented speech has lower quality than the original speech. However, when the fine-tuning is conducted on top of the pre-trained model, we obtain 8.7% relative WER reduction over the baseline model, which is comparable to the model that is finetuned on the multilingual seed model pre-trained with five languages including 75,000 hours labeled raw speech data. Finally, when we combine the augmented speech samples with the multilingual speech data to do model pre-training, another extra 3.8% relative WER reduction is achieved.

The rest of this paper is organized as follows. We discuss related work in Section 2. The Transformer transducer models used in all our experiments are briefly introduced in Section 3. The proposed speech data augmentation method is described in Section 4. Experiments are presented in Section 5. In Section 6, we conclude our work.

## 2. RELATED WORK

TTS is a popular way to synthesize speech from texts to generate speech data that can be used in speech model domain adaptation. For example, [20, 21] have used a multi-speaker neural TTS system to generate speech data using the text-only data of the new domain to adapt the RNN-T model. [22] has applied a TTS engine to synthesize audio data from the text for the target language to train the speech model. Since TTS-generated audios are typical of sub-optimal quality in terms of speaker and acoustic coverage as compared to real-world audios, in order to mitigate this effect, [21, 22] have proposed to freeze the encoder when applying augmented TTS data in model training. TTS generated speech data have been also used into the model pre-training as in [23]. However, TTS generated speech data cannot include the variations of real speakers and acoustic conditions that are very important in the model training, which usually leads to inferior model performance as in [21].

[24] is another data augmentation method that replaces the aligned audio representations with the predicted tokens from an audio dictionary that has been generated from existing transcribed data. The predicted tokens could be generated by a language model such that the augmented data pairs are semantically close to the original data. [21] has introduced another method to generate speech audios with spliced data for speech model domain adaptation that improves the baseline and the adaptation with the TTS data by 58.03% and 15.25% relative word error rate reduction, respectively. However, these methods did not work on large scale text data and did not provide solutions to handle OOV words.

## 3. TRANSFORMER TRANSDUCER

A neural transducer model [25, 26, 27] is the most popular streaming E2E model [20] which has three components: an acoustic encoder network, a label prediction network, and a joint network. Neural transducer models can use different types of models as encoders such as LSTMs in RNN-T models [25] and transformers [28, 29, 30] in Transformer transducer models [31, 32, 33]. Each Transformer block in the encoder network is constructed from a multi-head self-attention layer followed by a feed-forward layer. Specifically, we use the Transformer transducer model in [33] for low-latency and low-computation-cost streaming ASR. The loss function of transducer models is the negative log posterior of output target label $y$ given input acoustic feature $x$ and is defined as

$$L = -\log P(y|x) \qquad (1)$$

which is calculated by the forward-backward algorithm described in [25].

Figure 1: *Alignment between power spectrum and grapheme letters that can construct words and sentence pieces for the Italian utterance "Che tempo fa"*

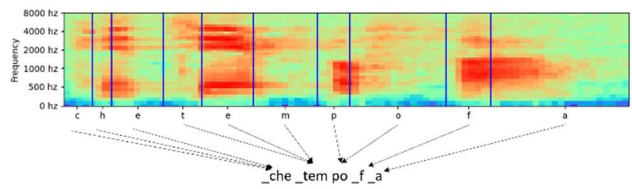

## 4. SPEECH DATA AUGMENTATION

### 4.1. Library establishment of augmented speech sample candidates

As senone-based models, hybrid chenone models can generate the alignment between speech features and grapheme letters [13]. The difference between senone and chenone models is the former is based on acoustic pronunciation while the latter is using grapheme letters directly. Figure 1 shows an example of alignment between speech power spectrum and grapheme letters generated by the chenone model for an Italian utterance "*Che tempo fa*". We build the libraries of augmented speech sample candidates with three levels of units as words, sentence pieces, and grapheme letters that can construct speech utterances for any texts without any OOV issue. All the boundary information for these three units can be obtained from the alignment between speech feature segments and grapheme letters as shown in Figure 1.

We first extract all feature segments of words in the ASR training data and build a list of all distinct words. The instances of feature segments of the same word are collected as the samples of this word. We also store a word feature segment index and its segment length into the library in order to easily access any word feature segments. The library building of sentence pieces and grapheme letters follows the same way as the word library building except that we need to generate the sentence pieces with a BPE model [34] for the speech transcription in advance. Since there could be huge amounts of instances available for some units in the training speech data, we limit the maximum number of words, sentence pieces, and grapheme letters in their libraries, and also filter out segment instances with unnormal durations. In addition, grapheme letters have the most feature segment instances. However, most of augmented speech data can be constructed by the word and sentence piece segments. Therefore, we keep an even smaller portion of grapheme feature segments that we will talk about in more detail in Section 5.

Figure 2: *Speech spectrum of the utterance "Che tempo fa" constructed by feature segment candidates from libraries of words, sentence pieces, and grapheme letters*

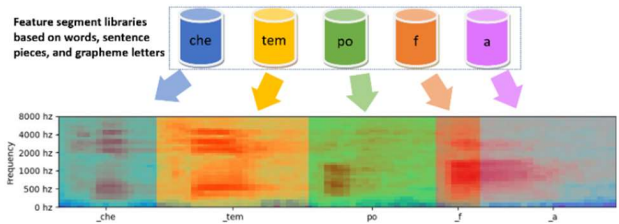

### 4.2. Speech feature utterance generation

We use text in LM training data to query the library we build in Section 4.1 to generate the new speech feature utterances. For each text sentence, we scan each single word, and if the word is in whole word list that we build in Section 4.1, we then randomly select a word instance of feature segments in this word library as the corresponding speech features for this word. If it is not in the word list, we check the availability of the word corresponding sentence pieces in the sentence piece list to decide if we can take feature segments from the sentence piece libraries. If a sentence piece that is composed of the word is not in the sentence piece list, we finally go to the grapheme letter libraries to extract grapheme feature segments. After we get all speech features corresponding to a text sentence from libraries of words, sentence pieces and grapheme letters, we concatenate all the feature segments to construct the whole speech feature utterance as shown in Figure 2. By using the three levels of units, we can generate speech features for any text.

Since all speech segment features are randomly selected from the libraries, there is no restriction that the selected feature segments have to be from the same utterance, the same speaker, or the same acoustic environment. This concatenation based on random speech feature selection can provide great flexibility to generate a variety of speech feature data. However, a concern for this method is it may break the nature of speech continuity and fluency. We would dismiss this concern with the following two aspects: 1) our concatenation method is based on speech filterbank features not on speech audios that are more affected by speech continuity. We also filter out the speech feature segments with abnormal duration length when we build up the speech feature segment libraries to guarantee the quality of feature segments. 2) We prioritize using larger units for concatenation and only fall back with smaller units when larger units are not available in dictionary. This helps to keep the concatenation as smooth as possible. Note that one property of E2E models is that they make prediction after processing a segment of speech instead of processing frame by frame. Therefore, this disfluency at the transitions between words or sentence pieces won't affect Transformer transducer models too much, compared to the benefit that the textual information injects into the model.

Our data augmentation method also has two advantages over the method of TTS-based audio generation as: 1) the cost is much lower since it doesn't need any extra TTS model to generate TTS audio. 2) The speech data generated by our method is closer to real speech at each segment than TTS-generated data since it can potentially cover all speakers and acoustic conditions in the real speech training data, and therefore the data variations are much more than that in TTS-generated data.

Another way to leverage text data is to initialize prediction network with a RNNLM trained with the text data [35]. However, the gain is very limited as reported in [35].

## 5. EXPERIMENTS

### 5.1. Experimental setups

*5.1.1. Data*
We evaluate the proposed method using an Italian ASR task. The training data covers multiple domain such as Dictation, Video, Cortana, Conversation, and Others that includes miscellaneous categories of data such as Xbox and Windows phone, etc. Similar to the training data, the test data also covers multiple domains. The data is completely anonymized. Overall, we have around 12,500 hours of raw speech training data and around 38 hours test speech data from mixed domains. As for the texts from LM training data, we randomly select 9 million sentences to generate the augmented speech feature utterances. Each text sentence includes around 10 to 20 words.

*5.1.2. Model structure and training configuration*
In our Transformer transducer models [33], 18 Transformer blocks with 320 hidden nodes, 8 attention heads, and 2048 feedforward nodes are used as the encoder; 2 LSTM layers with 1024-dimensional embedding and hidden layer are used in the prediction network. All our experiments use 80-dimensional filterbank features with 25 millisecond (ms) windows and 10ms shift. Two convolutional layers are applied to get features with 40ms sampling rate. The input acoustic feature sequence is segmented into chunks with a chunk size of 4 in our experiments and chunks are not overlapped. Therefore, the maximum lookahead is 160ms. In addition, we also apply 18 left frames to leverage history acoustic information in our training. The learning rate warmup strategy is the same as in [36]. All the models are trained from scratch and with mixed precision for efficient training. Around 4k sentence pieces [19] from all Italian speech transcribed text data are used as token units in all related experiments. All multilingual models have 10k output nodes in order to cover more tokens from multiple languages, and each mini-batch consists of utterances from all languages, sampled according to their natural training data distribution [37, 38]. All finetuning experiments are conducted by initializing the model's weights from the pre-trained models except for the output layers that are randomly initialized. Finally, for the purpose of fair comparisons, all finetuning experiments with the original 12,5000 hours Italian data use the same peak learning value and learn rate decay scheduler.

Figure 3: Averaged d*uration histogram of grapheme letters in words and sentence pieces from speech training data*

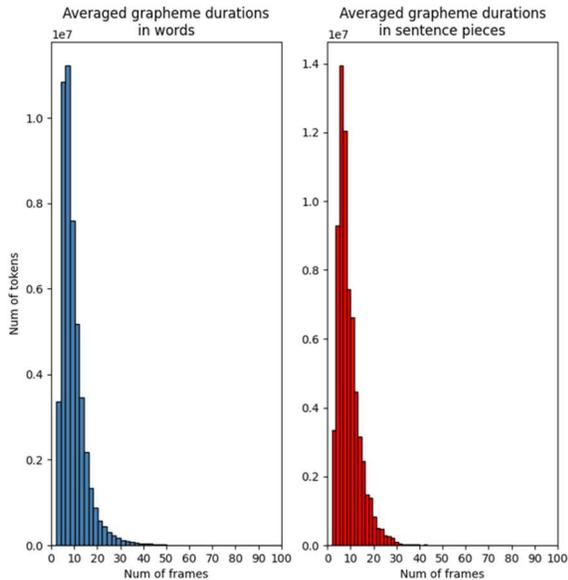

### 5.1.3. Distribution of feature segment duration

Since the grapheme letter duration generated from alignment varies a lot, we generate the statistics of feature segment durations based on grapheme letters that are used to construct the feature segments for words and sentence pieces. Figure 3 shows most of the averaged grapheme letters have durations that are less than 30 speech feature frames in both cases of word and sentence pieces. The most frequent average grapheme durations are between 10 and 20 speech feature frames while the max one can be 100 feature frames. In addition, silence as transitions between words should also have reasonable durations in speech utterances. Therefore, we limit the max averaged duration of 30 speech feature frames for a grapheme letter in words and sentence pieces, and 50 feature frames for silence between words or sentence pieces during the generation of speech feature utterances.

### 5.1.4. Unit selection strategy

As shown in [39], the whole word pieces can improve the E2E model performance in ASR. If the whole word pieces are not available, the sub-word pieces would be the second option [39]. We take the similar strategy for the unit selection in our method to generate speech feature utterances. We first try to find the whole word feature segments from the libraries with the highest priority. If it is not applicable, the second priority is to use sentence pieces, and if it is still not applicable, the last option is to use grapheme letters. From our statistics, for all 9 million text sentences, 70% of them can be covered by the whole word segments, and it is 99% coverage if we use both whole word and sentence piece segments. The grapheme letters are only needed to cover the remaining 1%. Therefore, we only keep 100 instances with reasonable duration length for each of grapheme letters in their data libraries since they are hardly used, while for the whole word and sentence piece, the maximum instance number is 500 for each unit. As we know, Italian is a language that the grapheme-to-phoneme mapping is mostly one-to-one, which may bring up extra benefit for our data augmentation method. However, since only 1% text data are covered by grapheme letters, Italian that happens to be a study case does not take much the advantage of one-to-one mapping between grapheme and phoneme and we believe the proposed idea can be well generalized to other languages.

Since the superior performance of a small-scale spliced data over TTS data was already reported for model adaptation purpose in [21], we don't do extra comparisons in our experiments between our data augmentation and TTS based methods because of the very large cost of TTS data generation for 9 million sentences. Another reason we don't compare with TTS data generation is it is not a popular and practical method to generate very large scale data for seed model training in transfer learning compared to using transcribed data from resource-rich languages to pre-train a seed model which is our main interest to compare in this paper.

## 5.2. Experimental results

### 5.2.1. Effect of cross-domain data generation

Since our speech training data is from different resources, we measure how the quality of generated speech feature utterances are affected between the cases of using the training data from a single domain and multi-cross domains. We randomly select 500,000 text sentences and generate around 1,000 hours speech feature utterance data by using the data from Dictation only domain and multi-cross domains including Dictation, Video, and Conversation. We train models based on the 1,000 hours data generated by these two different strategies plus another 1000 hours original transcribed data, and get the model evaluation results in Table 1. The model trained by using data generated from the single domain of Dictation achieve the WER of 28.66% that is much better than the model trained with multi-cross domain data with the WER of 32.18%, which indicates the better quality of augmented speech data generated from the same domain of speech training data. We have in total 5 domains of data as Dictation, Video, Cortana, Conversation, and Others. The domain Others includes very noisy data of different types of speech from different sub-categories. In addition, Cortana data has similar acoustic condition as Dictation but with much shorter utterance durations and simpler semantic context patters. Therefore, we finally remove the speech data from domains of Cortana and Others, and keep the data from Dictation, Video, and Conversation for the final speech data augmentation. We equally divide the total 9 million text sentences into three parts and each of them is generated by speech feature segments from one of the three domain training data. The total augmented speech data is 20,000 hours.

Table 1: *Evaluation results with models trained with original transcribed data and augmented data generated by single and multi-cross domain speech training data*

| Domain | WER |
|---|---|
| Single (dictation) | 28.66% |
| Multi-cross | 32.18% |

We also collect statistics about the distinct numbers of words, sentence pieces, and grapheme letters that are covered from the three domains of data in Table 2. Domain Video covers most words while Dictation has the least number of distinct words included. The distinct numbers of sentence pieces and grapheme letters are almost the same in these three domains of data.

Table 2: *Statistics of distinct occurrence from different units of feature segments in Video, Conversation, and Dictation data*

| Corpus | Word | Sentence Piece | Grapheme |
|---|---|---|---|
| Video | 215118 | 3536 | 28 |
| Conversation | 166561 | 3536 | 28 |
| Dictation | 99126 | 3532 | 28 |

*5.2.2. Apply augmented speech samples in model pre-training*

We evaluate models pre-trained and trained with different data assets. We first mix all augmented 20,000 hours speech data with the original transcribed 12,500 hours raw speech data for Italian to train a Transformer transducer model. As shown in Table 3, this model gets the WER of 14.59% that is 3.6% relative worse than the baseline Transformer transducer model with the WER of 14.08% that is trained with only original speech data. As an ablation study, we fix the original transcribed data and sample the augmentation data with different weights during training to measure its effect. However, we get either similar or worse WERs compared to the currently reported results. We then use the model trained with mixed data as the pre-trained model to finetune on top of it with only the original transcribed data and obtain the WER of 12.85% which is 8.7% relative WER reduction from the baseline model. We also build a multilingual seed model using transcribed 75,000 hours raw speech data from five languages as French, German, Italian, Portuguese, and Spanish, in which the original transcribed 12,500 hours Italian data is included, and the data amount from other languages is almost evenly distributed. We first evaluate this multilingual model on Italian test sets and get the WER of 14.36% that indicates this model is reasonably well trained. Based on the traditional transfer learning method, we then conduct the finetuning on the multilingual pre-trained model with only Italian data. This training gives similar WER of 12.87% to the model finetuned on the pre-trained model that is trained with only mixed 32,500 hours data, which implies, instead of applying transfer learning, less than half data augmented by our method can be used to pre-train a model that can achieve the similar model performance. In addition, consider the augmented data is also generated by the existing transcribed Italian data, there is no additional speech data required in this method while transfer learning needs lots of transcribed data from other languages. Finally, we merge the augmented 20,000 hours speech data with the multilingual 75,000 hours data to get a new pre-trained model, and then finetune with the original raw Italian speech data. Such training achieves the best WER as 12.36% which represents 12.2% relative WER reduction over the baseline model, which further verifies the effectiveness of our augmented data.

Table 3: *Evaluation results based on models pre-trained and finetuned by different training data*

| | WER |
|---|---|
| Baseline w/o pre-trained model | 14.08% |
| Model trained with mixed augmented and original transcribed data (1) | 14.59% |
| Finetuned model with (1) | 12.85% |
| Multilingual pre-trained model (2) | 14.36% |
| Finetuned model with (2) | 12.87% |
| Finetuned model with pre-trained model trained by multilingual + mixed data | 12.36% |

## 6. CONCLUSIONS

In this paper, we propose a new method to generate augmented speech feature utterances by using text from LM training data to query ASR training corpus. Our method has the advantage of low cost without using additional transcribed or unlabeled speech data. Compared to the TTS method, our method can integrate real multi-speaker and acoustic condition information into the generated data with very low cost. We mix 20,000 hours speech data generated by our method with the original transcribed speech data for model pre-training of Italian language, and achieve 8.7% relative WER reduction over the baseline model. Such model, with much smaller amount of training data, gives similar performance as the model transferred from a multilingual model trained with 75,000 hours raw speech data. When we merge the augmented speech data with the multilingual data to train a new seed model and then do the finetuning, we achieve further improvement with 12.2% relative WER reduction over the baseline.